\documentclass[regular]{jfm}

\usepackage{graphicx}
\usepackage{newtxtext}
\usepackage{newtxmath}
\usepackage{natbib}
\usepackage{hyperref}
\hypersetup{
    colorlinks = true,
    urlcolor   = blue,
    citecolor  = blue,
    linkcolor  = blue
}
\usepackage{layouts}

\shorttitle{Buoyancy-driven attraction of active droplets }
\shortauthor{Y.\,Chen and others}
\title{Buoyancy-driven attraction of active droplets}
\author{
Yibo Chen\aff{1},
Kai Leong Chong\aff{2}\corresp{\email{klchong@shu.edu.cn}},
Haoran Liu\aff{1 },
Roberto Verzicco\aff{1,3,4} \and
Detlef Lohse\aff{1,5}\corresp{\email{d.lohse@utwente.nl}}
}

\affiliation{
\aff{1}Physics of Fluids Group, Max Planck Center for Complex Fluid Dynamics and J.M.Burgers Center for Fluid Dynamics, University of Twente, P.O. Box 217, 7500 AE Enschede, The Netherlands
\aff{2}Shanghai Key Laboratory of Mechanics in Energy Engineering, Shanghai Institute of Applied Mathematics and Mechanics, School of Mechanics and Engineering Science, Shanghai University, Shanghai, 200072, PR China
\aff{3}Dipartimento di Ingegneria Industriale, University of Rome `Tor Vergata', Via del Politecnico 1, Roma 00133, Italy
\aff{4}Gran Sasso Science Institute - Viale F. Crispi, 7 67100 L'Aquila, Italy
\aff{5}Max Planck Institute for Dynamics and Self-Organisation, Am Fassberg 17, 37077 G\"ottingen, Germany
}

\begin{document}

\maketitle

\begin{abstract}
For dissolving active oil droplets in an ambient liquid, it is generally assumed that the Marangoni effect results in repulsive interactions, while the buoyancy effects caused by the density difference between the droplets, diffusing product and the ambient fluid are usually neglected. However, it has been observed in recent experiments that active droplets can form clusters due to buoyancy-driven convection (Kr\"uger \textit{et al.} \textit{Eur. Phys. J. E}, vol. 39, 2016, pp. 1-9). In this study, we numerically analyze the buoyancy effect, in addition to the propulsion caused by Marangoni flow (with its strength characterized by P\'eclet number $Pe$). The buoyancy effects have their origin in (i) the density difference between the droplet and the ambient liquid, which is characterized by Galileo number $Ga$, and (ii) the density difference between the diffusing product (i.e. filled micelles) and the ambient liquid, which can be quantified by a solutal Rayleigh number $Ra$. We analyze how the attracting and repulsing behaviour of neighbouring droplets depends on the control parameters $Pe$, $Ga$, and $Ra$. We find that while the Marangoni effect leads to the well-known repulsion between the interacting droplets, the buoyancy effect of the reaction product leads to buoyancy-driven attraction. At sufficiently large $Ra$, even collisions between the droplets can take place. Our study on the effect of $Ga$ further shows that with increasing $Ga$, the collision becomes delayed. Moreover, we derive that the attracting velocity of the droplets, which is characterized by a Reynolds number $Re_d$, is proportional to $Ra^{1/4}/( \ell /R)$, where $\ell/R$ is the distance between the neighbouring droplets normalized by the droplet radius. Finally, we numerically obtain the repulsive velocity of the droplets, characterized by a Reynolds number $Re_{\text{rep}}$, which is proportional to $PeRa^{-0.38}$. The balance of attractive and repulsive effect leads to $Pe\sim Ra^{0.63}$, which agrees well with the transition curve between the regimes with and without collision.

\end{abstract}


\section{Introduction}\label{sec:intro}

The fundamental principles of microorganisms propulsion have gained attention across disciplines over the past few decades \citep{brennen1977fluid, stone1996propulsion, lauga2009hydrodynamics, marchetti2013hydrodynamics, li2016collective, blackiston2021cellular}. Given the abundance of such microorganisms such as bacteria and plankton in our ecosystem \citep{hays2005climate}, studying their individual and collective motion is critical for understanding the dynamics of the entire ecosystem \citep{guasto2012fluid}. The interactions between microorganisms can be purely physical, i.e. based on hydrodynamics \citep{ramia1993role, ishikawa2006}, or biological, i.e. based on visual signals \citep{trushin2004light} or by chemical signals \citep{adler1975chemotaxis}. Disentangling these effects makes it difficult to analyze the various interactions in real microorganism colonies. To reduce the complexity, in recent years artificial microswimmers as a simplified model have been investigated in order to understand the interactions between living microorganisms \citep{pedley2016spherical, maass2016, datt2019active,  hokmabad2019, gompper2020, chen2021, li2022swimming}. Such artificial microswimmers are designed to propel themselves by converting free energy from the environment into kinetic energy \citep{ogrin2008}. Similar interactions as those between living microorganisms are observed, such as chemotaxis, collective entrainment, and cluster formation \citep{maass2016, lozano2016, jin2017, lohse2020, jin2021collective}. 

One extensively studied type of artificial microswimmer is a dissolving active oil droplet floating in water \citep{maass2016}. The driving mechanism behind the propulsion of such active droplets is the Marangoni effect. The basic feature is that whenever there is an inhomogeneity of surfactant concentration at the surface of the droplet, the consequent surface tension difference generates a tangential Marangoni flow adjacent to the surface, which leads to the self-propulsion of the droplet \citep{herminghaus2014, morozov2019, morozov2019self, michelin2022self}. This effect can also be generalised to other coupled systems such as the particles with catalytic surfaces. The resulting flows are then referred to as diffusio-phoretic flow \citep{anderson1989}. With the Marangoni effect or diffusiophoresis be present, such active droplets become repulsive. This simply happens because the Marangoni flow or the diffusiophoretic flow will propel the active droplet or particles towards higher surfactant concentration direction (where the surface tension is lower), and the concentration of surfactant molecules is lower between two close-by droplets or particles than that in the periphery.

Repulsive interactions induced by Marangoni effects between active droplets have been well studied in numerous experimental and theoretical works. A clear experimental observation of the repulsion was conducted by \cite{moerman2017solute}, who quantitatively measured the repulsive velocity for a pair of active droplets and analyzed the relations between the repulsive force and their distance. Later on, \cite{lippera2020} theoretically analyzed the repulsive interaction between a pair of droplets and identified different motion modes. In a further study, \cite{lippera2021alignment} investigated the repulsive interactions for a pair of obliquely-colliding droplets, and identified whether the droplets interact directly or through their chemical wake.  Besides the direct interaction between active droplets, \cite{jin2017} found the active droplets also show trail avoidance behavior. They reported that the active droplet emitted filled micelles in the wake play a role as chemical repellents and cause trajectory avoidance. Based on such observation, \cite{daftari2022self} developed a mathematical model to mimic the trail-avoidance with multiple active droplets. They observed that the active droplets could trap themselves due to the trail-avoidance, a feature which has been called transit self-caging behavior.

However, besides the Marangoni effect caused by the difference in surface tension, also buoyancy effect caused by the inhomogeneity in the density field can affect the flow, namely by natural convection. One interesting example of a droplet for which the interplay between Marangoni and buoyancy forces leads to rich dynamics is the phenomenon of the "jumping droplet" \citep{li2019bouncing, li2021marangoni, li2022marangoni}. In that case the droplet repeatedly jumps up due to the Marangoni effect, and then slowly sinks due to buoyancy \citep{li2019bouncing, li2022marangoni}. Another example is the evaporation of a binary micro-droplet, where buoyancy competes with Marangoni forces and can drive the convection inside the droplet, playing a crucial role in the evaporation process \citep{edwards2018density, li2018evaporation, li2020rayleigh, diddens2021competing}. Finally for a pair of fixed droplets, \cite{lopez2022oscillatory} reported an oscillatory flow near the droplet, which again is triggered by the competition between Marangoni and buoyancy effects.

Coming back to dissolving active droplet, also here, despite the repulsive interaction by the Marangoni effect, \cite{krueger2016} observed opposing collective behaviours when buoyancy is significant in the collective droplet system. They reported that the active droplets attract each other and form clusters hovering in the fluid and ascribed this finding to buoyancy effect because the collective behavior only occurs when the density difference between the solvent and the droplet is above a certain threshold. In a further study, \cite{vajdi2021spontaneously} investigated the spontaneous rotation of the cluster formed by the attracted droplets. However, an explanation of the detailed mechanism of the attraction is still missing, especially on how the buoyancy effect drives the attraction and overtakes the repulsion driven by the Marangoni effect. 

Inspired by the above-mentioned studies, we focus on the collective behavior of active droplets with buoyancy effects. Note that there are two different types of buoyancy effects, either by the density difference between the droplet and ambient fluid, or by the solutal density difference between the dissolving product (i.e. filled micelles) and the ambient fluid. Hereafter, we call them the "droplet buoyancy effect" and the "product buoyancy effect", respectively. In this work, we will quantitatively analyze the interplay between the Marangoni effect, the droplet buoyancy effect, and the product buoyancy effect. We first simulate the interaction between a pair of active droplets. The numerical simulations allow us to capture the flow field around the droplet and how it induces the droplet interaction. Then we develop a model to predict the attracting velocity based on the method of reflections and on Faxen's law. We test our model with systems of two and three interacting droplets. Then we analyze the repulsive velocity  (based on the Marangoni flow) from simulations of a pair of fixed droplets. Finally, by comparing these results with those from the attractive velocity model due to buoyancy effect, we obtain a good prediction for the regime transition of droplet collision.
 
The paper is organized as follows: We first describe the problem setup in Section \ref{sec2}. The numerical method and validations of the numerical scheme are provided in Section \ref{sec3}. We first qualitatively analyze the role of the diffusiophoretic effect (characterized by $Pe$) and the product buoyancy effect in Section \ref{sec_result} and then we analyze the role of the droplet buoyancy effect in Section \ref{sec4_2}. Next, we develop a model to explain the attraction and calculate the attracting velocity and how it scales in section \ref{sec6_1} and \ref{sec5_2} . The model is then further tested with the system of a pair of droplets and three droplets in section \ref{sec5_3}. Then the repulsive effect is analyzed with cases of a pair of fixed droplets in section \ref{sec7}. Finally, concluding remarks are given in section \ref{con}.

\begin{figure}
\centering
\centering \includegraphics[width=0.6\textwidth]{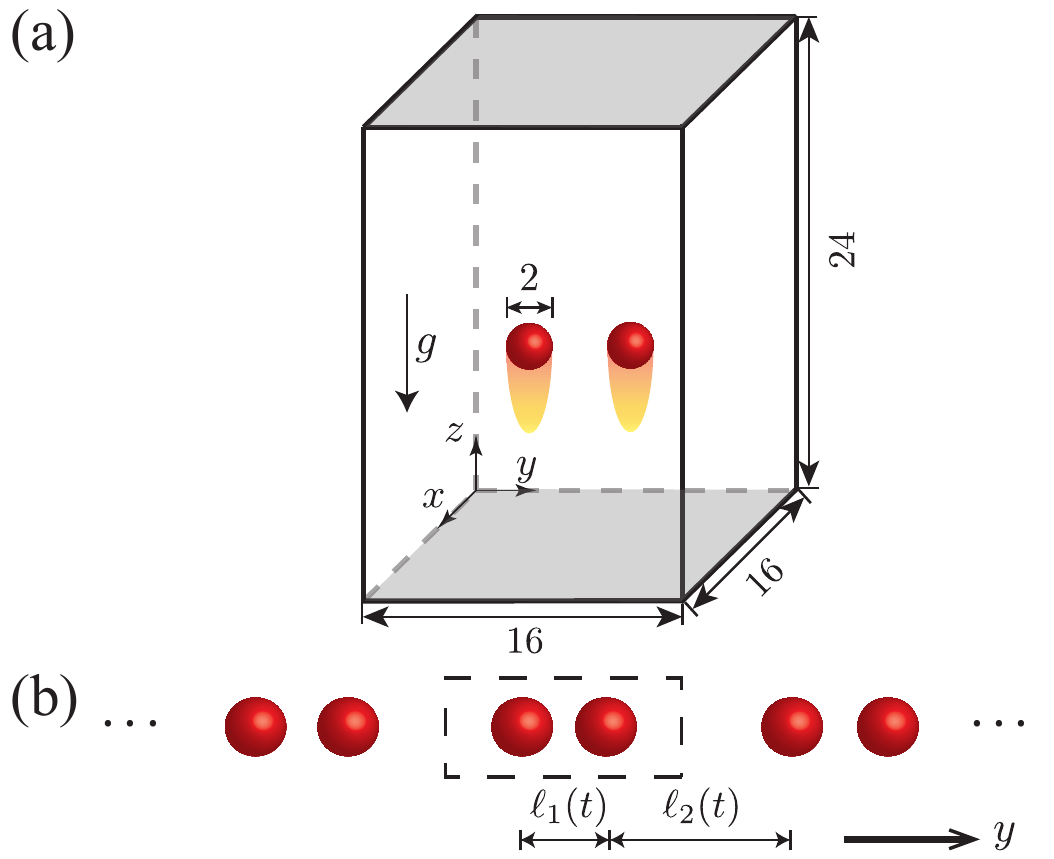}
\caption{The setup of the system. (a) The two droplets are initially located in the middle of the domain. The droplet buoyancy effect, the product (indicated by yellow tails under the droplets) buoyancy effect, and the diffusiophoretic effect are taken into consideration. The radius of the droplets is taken as characteristic length. The domain size expressed in this length is then $16\times 16 \times 24$. The numerical grid resolution is $161 \times 161 \times 241$. The top and bottom boundary conditions are set as solid wall (marked by gray plane) and the boundary conditions at $x$ and $y$ directions are periodic.  (b) Because of the periodic boundary condition in $y$ direction, the two droplets in the domain align with a series of droplets. In $x$ direction, the periodic boundary condition results in a balanced force. The distance between the two neighboring droplets inside the domain is $\ell_1$ and the distance of neighboring particles between inside and outside of the domain is $\ell_2$.}
\label{fig1}
\end{figure}

\section{Setup and control parameters} \label{sec2}

We start with a pair of active droplets in the surfactant solution sketched in figure \ref{fig1}. The gradual solubilization of the oil into the surfactant micelles causes a repulsive interaction via the Marangoni effect \citep{jin2017}. Simultaneously, the oil-filled micelles are generated near the droplet surface. Besides that, the droplet buoyancy effect and product buoyancy effect are taken into consideration in the simulations.

Considering similarities between diffusiophoresis and the Marangoni effect \citep{desai2021}, for simplicity, we focus on the phoretic effect induced by the concentration gradient of the filled micelles and will use the corresponding terminology. The physical variables to describe the system are the solutal concentration $\hat{c}$ and the velocity $\hat{\bf{u}}$. Note that all dimensional physical fields are marked with a hat (e.g. $\hat{c}$, $\hat{\bf{u}}$), while the dimensionless ones are without a hat (e.g. $c$, $\bf{u}$).

The droplets emit a solute (filled micelles) at a rate $\alpha >0$. The concentration boundary condition at the droplet surface is given by
\begin{equation} \label{eqj_alpha}
D\frac{\partial {\hat{c}}}{\partial {\hat{n}}}=-\alpha,
\end{equation}
where $D$ is the diffusion coefficient of the dissolution product, $\alpha$ the dissolution rate at the surface, and $\partial \hat{c}/\partial \hat{n}$ the concentration gradient normal to the surface.

The tangential concentration gradient at the surface induces a slip velocity, which is the so-called diffusiophoretic flow. The magnitude of the slip velocity $u_s$ is proportional to the local tangential concentration gradient, given by
\begin{equation} \label{eqj_us}
\hat{u}_s=M\nabla_s{\hat{c}},
\end{equation}
where $M$ is the mobility and $\nabla_s$ represents the tangential gradient. Since the filled micelles are emitted as a chemical repellent, the case of $M>0$ is considered.

We define $\hat{\rho}_0$ as the density of the surrounding fluid without any dissolved product and $\hat{\rho}_d$ as the density of the droplet itself. Note that in the experiments by \cite{krueger2016}, the density difference among $\hat{\rho}_0$, $\hat{\rho}_d$ and the density of the dissolving product (filled micelles) $\hat{\rho}$ is lower than $3\%$. Therefore, we consider the density difference within the Boussinesq approximation, i.e. the density of the fluid $\hat{\rho}$ is assumed to be linearly proportional to the filled micelle concentration
\begin{equation} \label{eqj_rho}
\hat{\rho}({\hat{c}})=\hat{\rho}_0(1+\beta \hat{c}),
\end{equation}
where $\beta$ is the proportionality constant between the density and the product concentration.

The velocity field outside the droplets is governed by the Navier-Stokes equations and the product concentration field by the advection-diffusion equation. The equations are non-dimensionalized by $R$ for lengths, $c_0=\alpha R/D$ for concentrations, and $\alpha M /D$ for velocities. Then the non-dimensional governing equations can be written as
\begin{equation}\label{eqj2}
\frac{\partial c}{\partial t}+\boldsymbol{u}\cdot\nabla c= \frac{1}{Pe}\nabla^2 c,
\end{equation}
\begin{subeqnarray}\label{eqj4}
\gdef\thesubequation{\theequation \mbox{\textit{a}},\textit{b}}
\frac{\partial \boldsymbol{u}}{\partial t}+(\boldsymbol{u}\cdot\nabla)\boldsymbol{u}=-\nabla p+\frac{Sc}{Pe}\nabla^2 \boldsymbol{u}-\frac{RaSc}{Pe^2}c\boldsymbol{e}_z,\quad
\boldsymbol{\nabla} \cdot \boldsymbol{u}=0,
\end{subeqnarray}
\returnthesubequation
and the velocity of the droplet $\boldsymbol{U}_d$ satisfies
\begin{equation} \label{dropvel}
\frac{d\boldsymbol{U}_d}{dt} = \frac{3Sc}{4\pi Pe}\int (\boldsymbol{\tau} \cdot \boldsymbol{n})dS-\frac{Ga^2Sc^2}{Pe^2}\boldsymbol{e}_z,
\end{equation}
where $\int (\tau \cdot n) dS$ is the force integrated over the surface of the droplet and $\boldsymbol{e}_z$ is the unit vector of the $z$ axis.

The dimensionless control parameters of these equations are the Rayleigh number $Ra$, which represents the strength of the product buoyancy effect,
\begin{equation} \label{Rayleigh}
Ra=\frac{c_0 \beta R^3 g}{\nu D} \text,
\end{equation}
the P\'eclet number $Pe$, which indicates the strength of the diffusiophoretic effect,
\begin{equation} \label{Peclet}
Pe=\frac{\alpha M R}{D^2},
\end{equation}
the Schmidt number $Sc$
\begin{equation} \label{Sc}
Sc=\frac{\nu}{D},
\end{equation}
which is kept as a constant in our study, and the Galileo number $Ga$, which represents the strength of the droplet buoyancy effect,
\begin{equation} \label{Ga}
Ga=\frac{\sqrt{|\hat{\rho}_d/\hat{\rho}_0-1|gR^3}}{\nu}.
\end{equation}
The concentration boundary condition (equation (\ref{eqj_alpha})) at the droplet surface reads in non-dimensional form
\begin{equation} \label{concengrad}
\frac{\partial {c}}{\partial {n}}=-1.
\end{equation}
The non-dimensional version of the velocity boundary condition at the droplet surface (equation (\ref{eqj_us})) is
\begin{equation}\label{eqjus}
u_s=\nabla_s c.
\end{equation}
We apply periodic boundary conditions along the horizontal directions ($x$ and $y$) of the domain (figure \ref{fig1} (b)), and solid wall boundary condition at the top and bottom of the domain. The concentration and velocity boundary conditions at the top and bottom are, respectively,
\begin{equation}\label{eqj5}
\frac{\partial c}{\partial z}=0,
\end{equation}
and
\begin{equation}\label{eqj5}
\boldsymbol{u}=0.
\end{equation}

\section{Numerical methods and validation} \label{sec3}

The Navier-Stokes equations and advection-diffusion equation are solved using direct numerical simulation (DNS) in Cartesian coordinates. We spatially discretize the equations using the central second-order finite difference scheme. Uniform staggered grids are used in all directions. The time integration is accomplished by using a fractional-step method. The non-linear terms are computed explicitly by a low-storage third-order Runge-Kutta scheme and the viscous and diffusion terms by a Crank-Nicolson scheme \citep{verzicco1996,van2015,ostilla2015multiple,spandan2018fast}. The model for the droplet-droplet and droplet-wall collision is based on the spring-dashpot model by \cite{costa2015}.

Because the vicinity of the surface is adopted to satisfy the constant normal fluxes and slip velocity boundary condition, we cannot allow the gap between the droplets to reach zero. Therefore, when the gap width is below 2 grid spacings, we assume that the droplets are in contact.

The numerical setup is shown in figure \ref{fig1}. The radius $R$ of the droplet is the characteristic length of the system, and the domain size is $L_x\times L_y\times L_z =16 \times 16 \times 24$. Two droplets of unit radius are initially aligned along the $y$ axis at the center of the domain with an initial distance $L_0=4$. We use uniform grids $N_x\times N_y\times N_z=161\times 161\times 241$. Since there is the periodic boundary condition along the $y$ axis, the simulations with two droplets are actually a part of a series of droplets aligning along the $y$ axis.  In $x$ direction, the periodic boundary conditions result in a balanced force. We use $\ell_1(t)$ and $\ell_2(t)$ to denote the distance between the droplet and its two neighboring droplets along the $y$ axis, and we further define $\ell (t)=\text{min}(\ell_1(t), \ell_2(t))$ as the droplet's distance to its nearest neighbor as shown in figure \ref{fig1} (b). 

We will take a range of parameters based on the data in the experiment by \cite{krueger2016}: P\'eclet number $0.5 \le Pe \le10$, Rayleigh number over the range $0.1\leq  Ra\leq245$ and Galileo number $0 \le Ga \le0.19$. The Schmidt number in the experiments is at the order of $10^4$. However, the simulations at such high $Sc$ are challenging due to the very small diffusivity compared to viscosity. Therefore, in our study, we set the Schmidt number $Sc=100$.

\begin{figure}
\centering
\centering \includegraphics[width=0.6\textwidth]{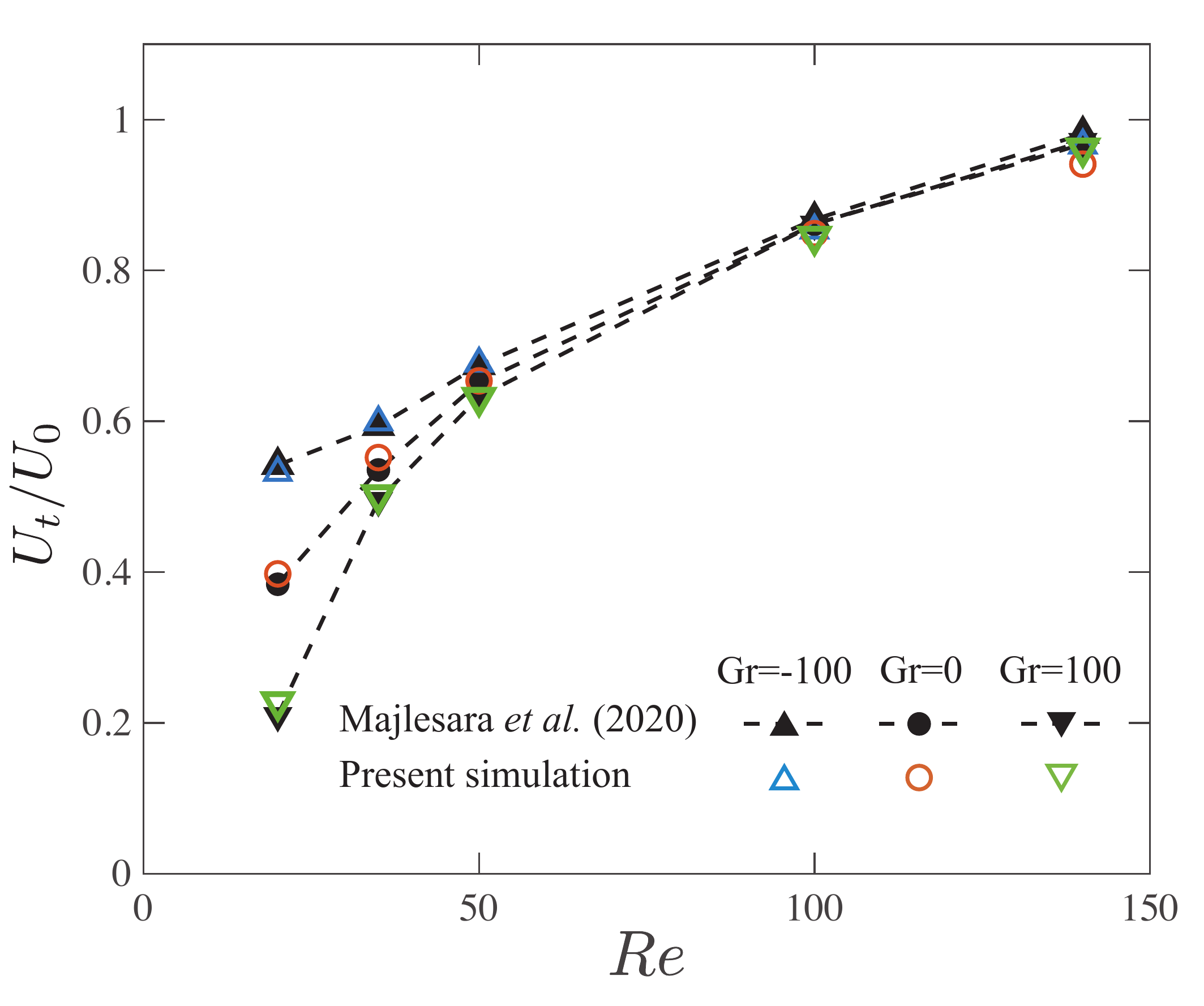}
\caption{Code validation for a settling particle at fixed temperature in a long vertical channel. The terminal velocity $U_t$, normalized by the reference velocity $U_0$, versus the Reynolds number $Re$, which is linearly correlated to Galileo number, $Re=\frac{2\sqrt{3}}{3}Ga$. We show results for three Grashof numbers $Gr=Ra/Pr$. The results obtained by \cite{majlesara2020}  are indicated by filled symbols with the dashed lines. Our simulations are represented by the opened symbols, showing excellent agreement.}
\label{fig2}
\end{figure}

Our code has been used to simulate diffusiophoretic particles. For the corresponding code validation, we refer the readers to our previous work \citep{chen2021}. As a further validation, we test our code by simulating  particle-laden flow with both droplet buoyancy effect and product buoyancy effect, and comparing with the existing results from the literature \citep{majlesara2020}.  These authors consider the cases about sedimenting cold/hot (fixed temperature) spherical particles in a long vertical fluid channel and study their terminal velocity. In that work, the product buoyancy effect is induced by the temperature variation, characterized by the Grashof number $Gr=Ra/Pr$. The particle buoyancy effect is characterized by the Reynolds number $Re$, which carries the same information as the Galileo number, $Re=\frac{2\sqrt{3}}{3}Ga$. We apply our code to simulate the same cases as \cite{majlesara2020}, and compare the normalized terminal velocity $U_t/U_0$ for various $Re$ and $Gr$, where $U_0$ is the characteristic buoyancy velocity. The numerical results are plotted in figure \ref{fig2}; they agree very well with those by \cite{majlesara2020}. 

\begin{figure}
\centering
\centering \includegraphics[width=1.0\textwidth]{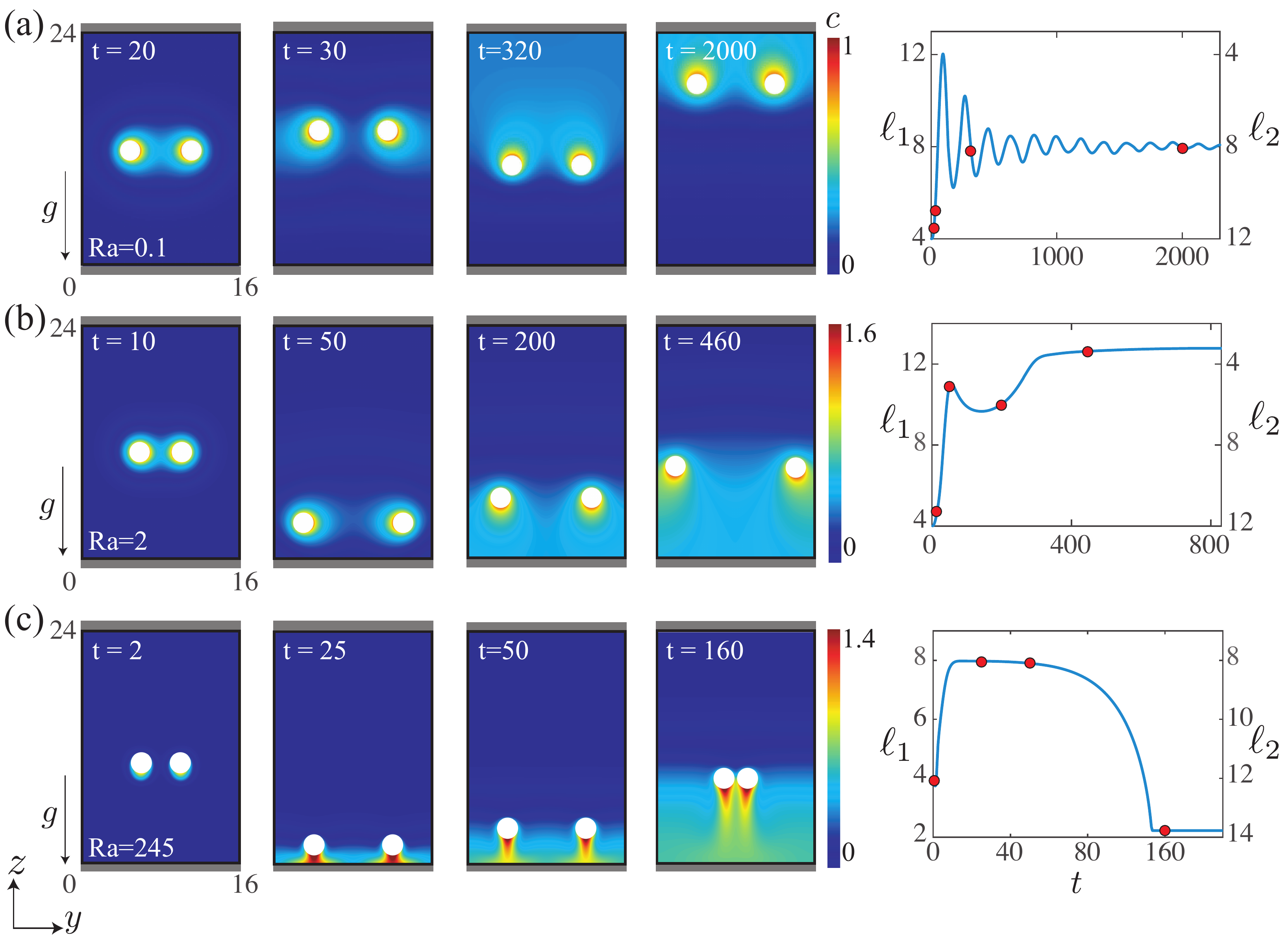}
\caption{Concentration contours (left panels) and distance between droplets (right panels) as funtion of time for a pair of droplets in the domain with parameters $Ga=0$, $Sc=100$, $Pe=5$ and various $Ra=$ $0.1$ (a), $Ra=2$ (b) , $Ra=245$ (c). At the right we plot the distances $\ell_1$ and $\ell_2$ defined in figure \ref{fig1} as a function of time. The droplet distances corresponding to the concentration contours are indicated as red filled circles in the plots.}
\label{fig3}
\end{figure}

\begin{figure}
\centering
\centering \includegraphics[width=1.0\textwidth]{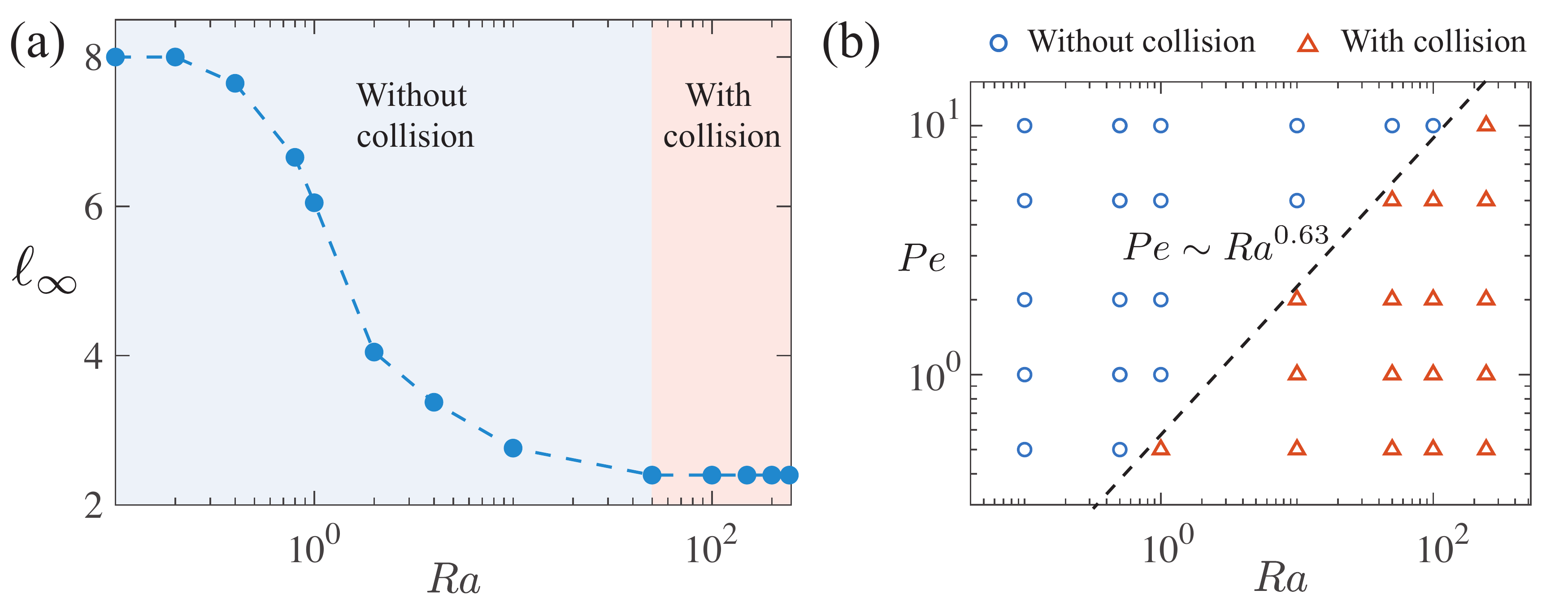}
\caption{ (a) Terminal distance $\ell_\infty$ between the nearest droplets with $Ga=0$, $Sc=100$, $Pe=5$ and different $Ra$ from $0.1$ to $245$. The dashed curve is a guide to the eyes. Two interaction modes are identified, marked with different colors: $Ra \leq 50$, the droplets remain at an equilibrium distance (without collision: blue), $Ra \geq 50$, the droplets collide with each other due to the strong attraction (with collision: red). (b) The interaction modes for $Ga=0$, $Sc=100$, $0.5 \le Pe \le 10$, $0.1 \le Ra \le 245$. The blue circles represent the cases without collision, while the red triangles those with collision. The results indicate that a higher Pe results in a higher Ra threshold, above which the collision occurs. }
\label{fig4}
\end{figure}

\begin{figure}
\centering
\centering \includegraphics[width=0.75\textwidth]{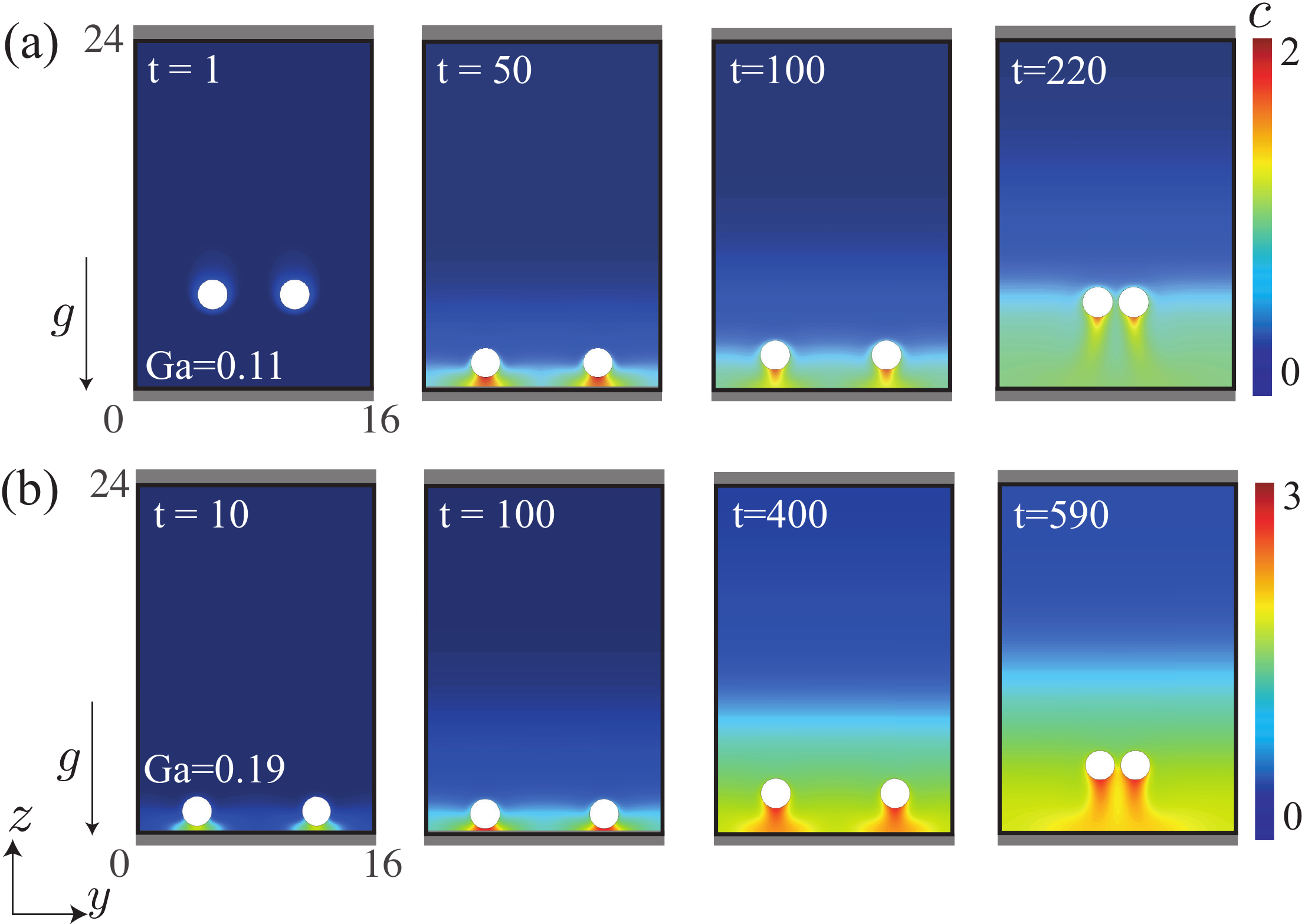}
\caption{Concentration contours for a pair of droplets with $Sc=100$, $Pe=5$, $Ra=245$ and two different $Ga$, (a) $Ga=0.11$ (b) $Ga=0.19$.
}
\label{fig5Ga_con}
\end{figure}

\section{Effect of P\'eclet number and of Rayleigh number}\label{sec_result}

In this section, we first investigate the role of the P\'eclet number and of the Rayleigh number by simulating the interaction between a pair of droplets with $Sc=100$, $0.5 \le Pe \le 10$, and $0.1 \le Ra \le 245$. The droplet buoyancy effect is absent in this subsection ($Ga=0$), and will be analyzed in subsection \ref{sec4_2}.

To demonstrate different interaction modes, we first focus on the cases $Pe=5$ and $Ra= 0.1$, $2$ and $245$ in figure \ref{fig3}, all for $Sc=100$ as throughout in this paper. For $Ra=0.1$, the diffusiophoretic effect is dominant. The mutual repulsion drives the droplets to the horizontally balanced positions ($1/4L_y$ and $3/4L_y$). The repulsion of the neighbouring droplets acts as restoring force to the balanced position, while the droplets also experience a damping force due to the viscous drag. Therefore the droplets perform a damped oscillation in horizontal direction around the balanced positions. In vertical direction, the droplets rebound from the walls because of the concentration accumulation in between.

As $Ra$ increases to $2$, the droplets approach their neighoring droplets. The reason is that the equi-distance balanced position becomes an unstable equilibrium due to the attraction between the droplets. In the end, the two droplets reach a new balanced point of finite distance $\ell<L_y/2$. Along the vertical direction ($z$), due to the stronger product buoyancy effect, the droplets first sediment to the bottom. Then the droplets gradually float up as the concentration between the wall and the droplets accumulate. As $Ra$ further increases, the terminal distance between the neighbouring droplets decreases. For $Ra=245$, the product buoyancy effect becomes even stronger. The droplets are more attractive to each other and collide in the end.

From the results, we find that different strengths of the product buoyancy effect lead to different terminal distances between the droplets along the horizontal axis. Therefore, we define the terminal distances $\ell_\infty$ to quantify that strength:
\begin{equation}\label{eqjtmdis}
\ell_\infty =\underset{t \to \infty}{\text{lim}} \ell(t) = \underset{t \to \infty}{\text{lim}} (\text{min}(\ell_1(t), \ell_2(t))).
\end{equation}
The dependence of $\ell_\infty$ on the Rayleigh number $Ra$ is shown in figure \ref{fig4} (a). We identify two different types of interaction according to $\ell_\infty$: (a) $Ra < 50$: without collision, where the droplets remain at an equilibrium distance without colliding with each other. The distance $\ell_\infty$ between the droplets reduces as $Ra$ increases. (b) $Ra \geq 50$, with collision, where the droplets collide due to the sufficiently strong attraction driven by the product buoyancy effect. 

We also simulate cases for different $Pe$ and $Ra$. The results can be classified into the two mentioned interaction modes, which are presented by different symbols in figure \ref{fig4} (b). The results indicate that there is competition between repulsion by diffusiophoresis and attraction by the product buoyancy effect. Higher $Pe$ results in a higher $Ra$ threshold, above which the collision occurs. This complies with the experimental results by \cite{krueger2016}, who find that the surfactant concentration ($Pe$) is increased, higher density differences ($Ra$) are needed for collective behavior to occur.

In summary, we numerically observe the interaction between droplets. We find very similar features as in the experiments by \cite{krueger2016}. While the diffusiophoretic effect (characterised by $Pe$) results in repulsion between droplets, the product buoyancy effect (characterised by $Ra$) leads to their attraction. We identify two different interaction modes: when the diffusiophoretic effect is dominant, the droplets reach a finite distance without collision; when the product buoyancy effect is dominant, the droplets collide in the end. In the next section, we will further investigate the role of the droplet buoyancy effect.

\section{Effect of increasing Galileo number} \label{sec4_2}
In this section, we analyze the influence of the droplet buoyancy effect, as quantified by the Galileo number $Ga$. We numerically investigate the interactions between a pair of droplets with $Pe = 5$, $Ra = 245$ and $0 \le Ga \le 0.19$. 

Snapshots at different times are plotted in figure \ref{fig5Ga_con}. For all examined cases with different $Ga$, the droplets collide in the end. To further analyze the interaction, we will examine the temporal change of the horizontal distance $\ell$ and the vertical height of the droplets.

\begin{figure}
\centering
\centering \includegraphics[width=0.95\textwidth]{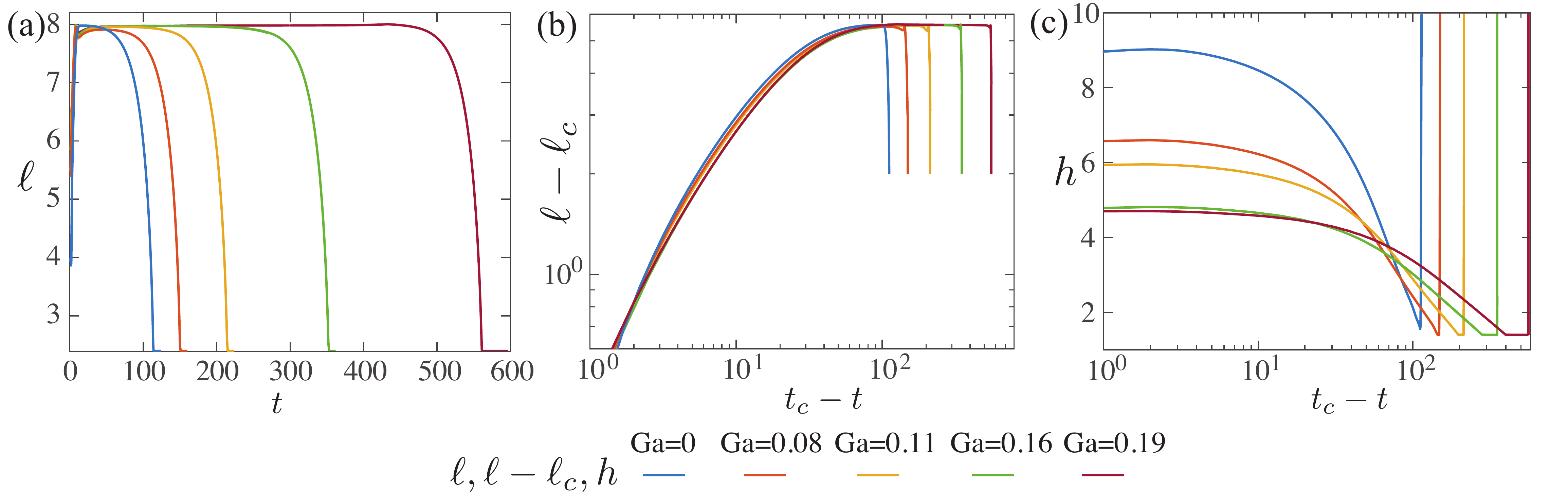}
\caption{The plot of distance $\ell$, $\ell-\ell_c$and height $h$ versus time $t$ or $t_c-t$ with $Sc=100$, $Pe=5$, $Ra=245$ and different $Ga$. $t_c$ and $\ell_c$ are the collision time and distance. (a) The distance between the two droplets $\ell$ as a function of time for different $Ga$. (b) $\ell-\ell_c$ and (c) $h$ along time $t_c-t$, where $\ell_c$ is the distance at collision point and $t_c$ is the collision time.
}
\label{fig5lh}
\end{figure}

We first plot the horizontal distance $\ell$ versus time $t$ in figure \ref{fig5lh} (a). The plot indicates that as $Ga$ increases, the waiting time for the collision to occur is longer. Next, we have a close inspection of the movement of droplets near the moment of collision through plotting $\ell-\ell_c$ as a function of $t_c - t$  in log-log scale in figure \ref{fig5lh} (b),  where $\ell_c$ is the collision distance and $t_c$ is the collision time. Remarkably, all curves collapse on each other near the collision point. It suggests that the attracting behavior of the droplets are mainly determined by $Ra$ and $Pe$, while the change of $Ga$ only leads to a delayed collision.

We also plot the height $h$ (right y axis) along $t_c-t$ in figure \ref{fig5lh} (c). As $Ga$ increases, the droplets wait for longer time before the occurrence of the approaching stage, and the rising velocity is also smaller for larger $Ga$. This is because a longer time is needed to build up a sufficiently large vertical concentration gradient to lift up a heavier droplet.

\begin{figure}
\centering
\centering \includegraphics[width=0.85\textwidth]{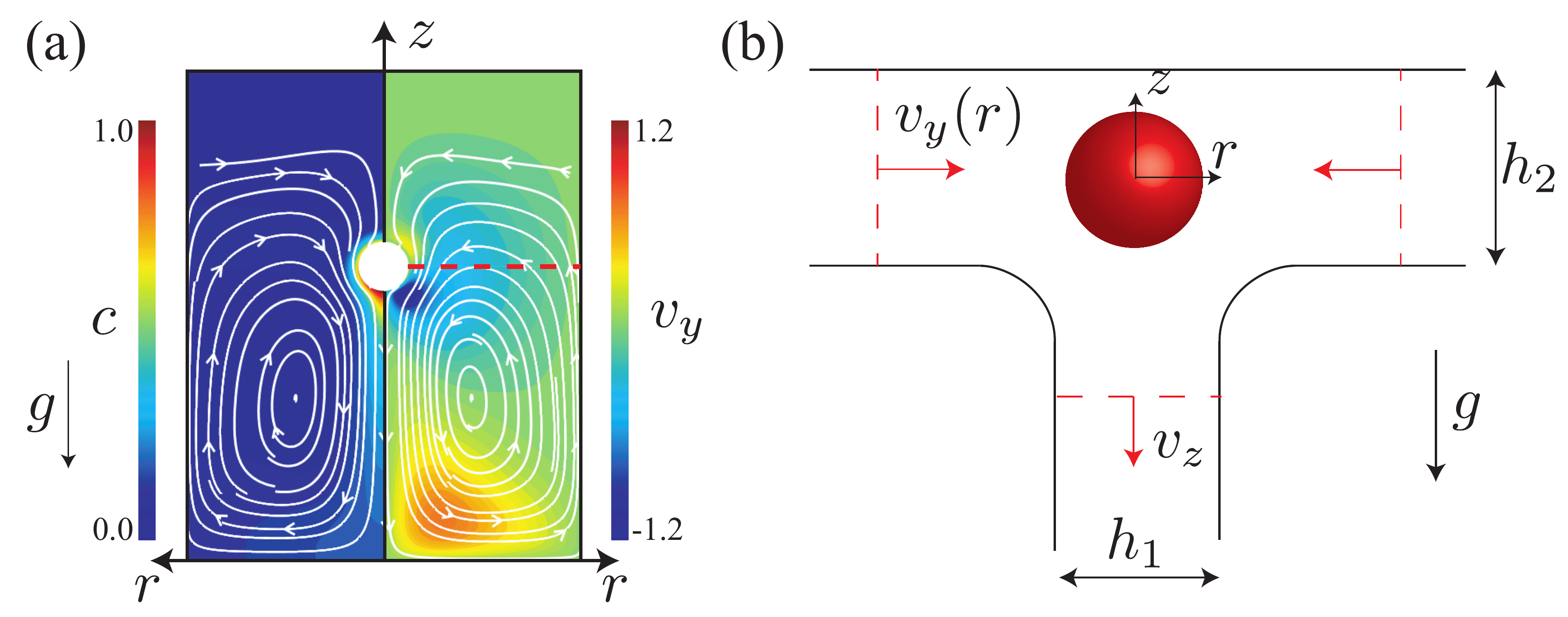}
\caption{(a) Concentration (left half) and velocity (right half) fields near a single droplet at $Ra=245$. The streamlines are shown by the white curves. The red dashed line is at the same height as the droplet. (b) The symmetric model is plotted in cylindrical coordinate $(r, z)$ to describes the flow near the droplet with buoyancy. The buoyancy induces a strong downwards flow under the droplet and a horizontal flow near the droplet. The width of the downwards flow is $h_1$ and the horizontal one $h_2$. In the simulation, we define the width $h_1, h_2$ of each flow branch by the width between $10\%$ of the maximum vertical and horizontal velocity.}
\label{fig5}
\end{figure}

\begin{figure}
\centering
\centering \includegraphics[width=0.5\textwidth]{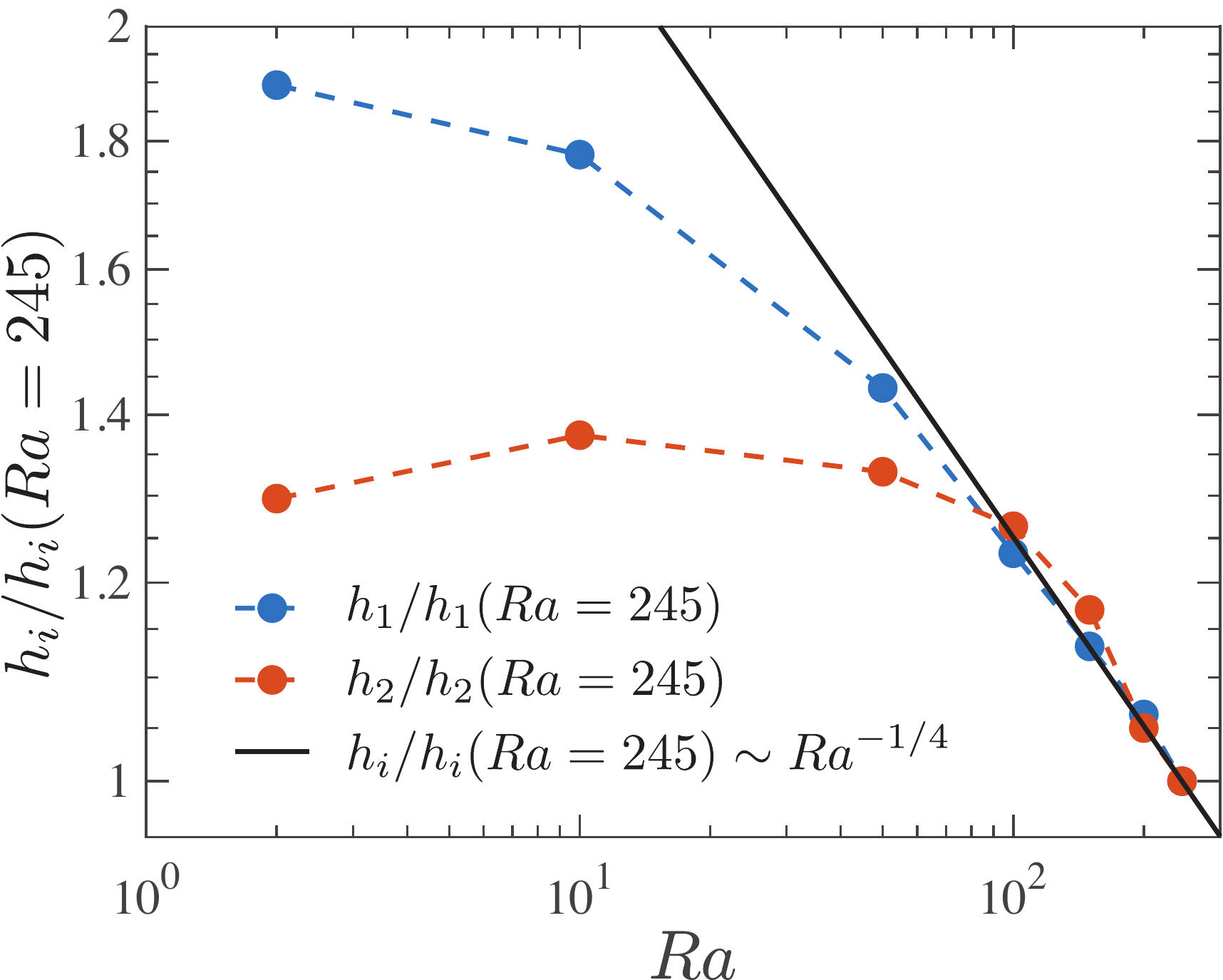}
\caption{The width of the downwards flow ($h_1$) and the horizontal flow ($h_2$) normalized by the corresponding height at $Ra=245$ for different $Ra$. The blue and red symbols are correspondingly the numerical results for $h_1$ and $h_2$. The solid curve represents (\ref{eqj4_1}).}
\label{fig6}
\end{figure}

\section{Attraction model with buoyancy} \label{sec_model}
In this section, we further investigate the origin of attraction and develop a model to estimate the attracting velocity in the buoyancy-dominant cases. Employing the point heat source model, we first derive a scaling law for the horizontal velocity around a droplet, and then calculate the attractive velocity, using the methods of reflections and Faxen's law. Since the droplet buoyancy effect only leads to delayed collision, we neglect it, i.e. we assume $Ga=0$ throughout this section.

\subsection{The velocity field near a single droplet} \label{sec6_1}

We start by simulating a single droplet to investigate the flow around it. Figure \ref{fig5} (a) shows the concentration and horizontal velocity ($v_y$) around a single droplet at $Ra=245$. From the fields, we observe a strong downwards plume, which leads to a higher concentration underneath the droplet. In the meantime, a horizontal flow is induced sidewards of the droplet. This inward flow drives the attraction between two nearby droplets.

We represent the buoyancy-driven flow near a single droplet by the schematics in figure \ref{fig5} (b). Since the flow around a single droplet is axisymmetric, it is illustrated in cylindrical coordinates $(r, z)$. There is horizontal inward flow sidewards of the droplet, and vertical downwards flow under the droplet. 

A similar case that has been well studied is the natural convection near a heat source or dissolutions source \citep{fujii1963,  moses1993experimental, dietrich2016}. \cite{fujii1963} theoretically studied the buoyancy-driven convection near a fixed heat source in the fluid, and quantitatively obtained  the buoyancy driven velocity. The theoretical results were later verified in experiments with a heating sphere in a fluid by \cite{moses1993experimental}. Both the velocity and the width of the plume scale with the Rayleigh number $Ra$ \citep{fujii1963,  moses1993experimental, dietrich2016}:
\begin{equation}\label{eqj4_0}
v_z\sim \frac{D}{R}Ra^{1/2},
\end{equation}
\begin{equation}\label{eqj4_1}
h/R \sim Ra^{-1/4}.
\end{equation}

We define the width $h_1, h_2$ of each flow branch as the distance between $10 \%$ of the maximum vertical and horizontal velocity. Due to the limited domain size, the vertical and horizontal flow cannot attain the asymptotic velocity of 0, preventing us from using a smaller threshold for the $h_1, h_2$ definition. The normalized values $h_1/h_1(Ra=245)$ and $h_2/h_2(Ra=245)$ versus $Ra$ obtained in simulations are plotted in figure \ref{fig6}. When $Ra$ is large enough ($Ra \geq 100$), the width of the channel well agrees with (\ref{eqj4_1}). For $Ra < 100$, the numerical results deviate, because of the existence of a strong enough diffusiophoretic effect.

Given the width of both the horizontal and the vertical flow, by continuity, we can further derive the relationship between the strength of the two velocities ($v_y$ for horizontal and $v_z$ for vertical), namely
\begin{equation}\label{eqj4_2}
v_y(r) \times 2\pi rh_2\sim v_z\times \pi h_1 ^2/4,
\end{equation}
where $r$ refers to the horizontal distance from the droplet center (along the red dashed curve in figure \ref{fig5} (a)). Then we define the local Reynolds number $Re_y (r)$ using the horizontal velocity $v_y (r)$. With (\ref{eqj4_0}) and (\ref{eqj4_2}), we obtain:
\begin{equation}\label{eqj4_3}
Re_y (r)=\frac{v_y(r) R}{\nu}\sim \frac{1}{Sc} \frac{Ra^{1/4}}{r/R} .
\end{equation} 

We verify (\ref{eqj4_3}) with the numerical simulations of a single droplet in the domain. Note that due to the periodic boundary condition, the horizontal velocity is also influenced by the neighboring droplets outside the domain,
\begin{equation}\label{eqj4_4}
\frac{Re_y (r)}{Ra^{1/4}}\sim \frac{R}{r}-\frac{R}{L_y-r}.
\end{equation} 

\begin{figure}
\centering
\centering \includegraphics[width=0.5\textwidth]{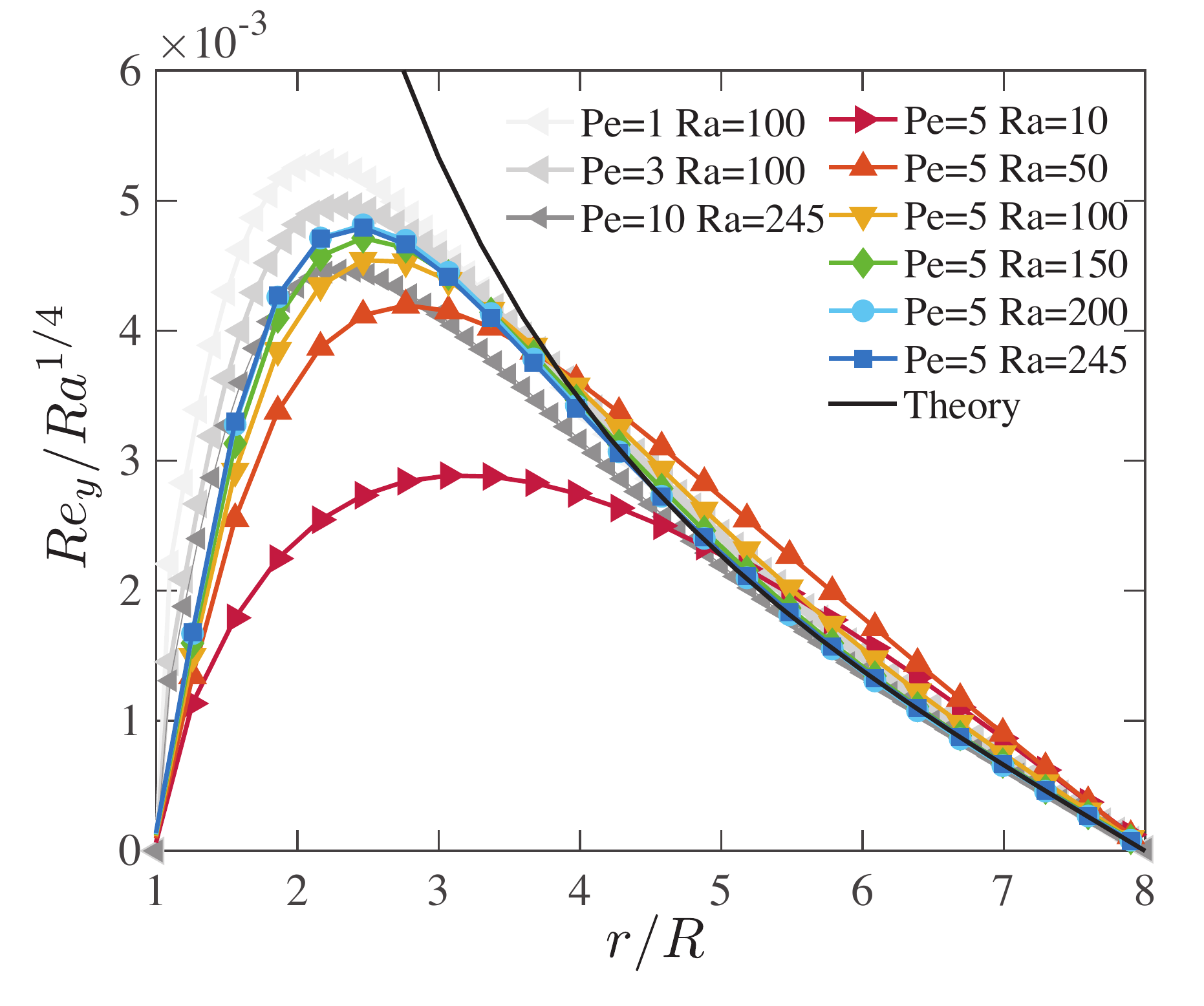}
\caption{$Re_y(r)$ normalized by $Ra^{1/4}$ along the red dashed line in figure \ref{fig5}(a) for various distances $r$ to the droplet center normalized by the radius $R$ of the droplet. The markers are the numerical results and the solid lines are a guide to the eye. The solid black curve represents relationship (\ref{eqj4_4}) with a fitted prefactor $0.021$. }
\label{fig7}
\end{figure}

The results for different $Ra$ and $Pe$ are shown in figure \ref{fig7}. The numerical results agree well with the theory equation (\ref{eqj4_4}) for $r/R>4$. The results deviate near the droplet surface $r/R<4$, because the horizontal velocity reduces to zero approaching the droplet surface. 

\begin{figure}
\centering
\centering \includegraphics[width=0.5\textwidth]{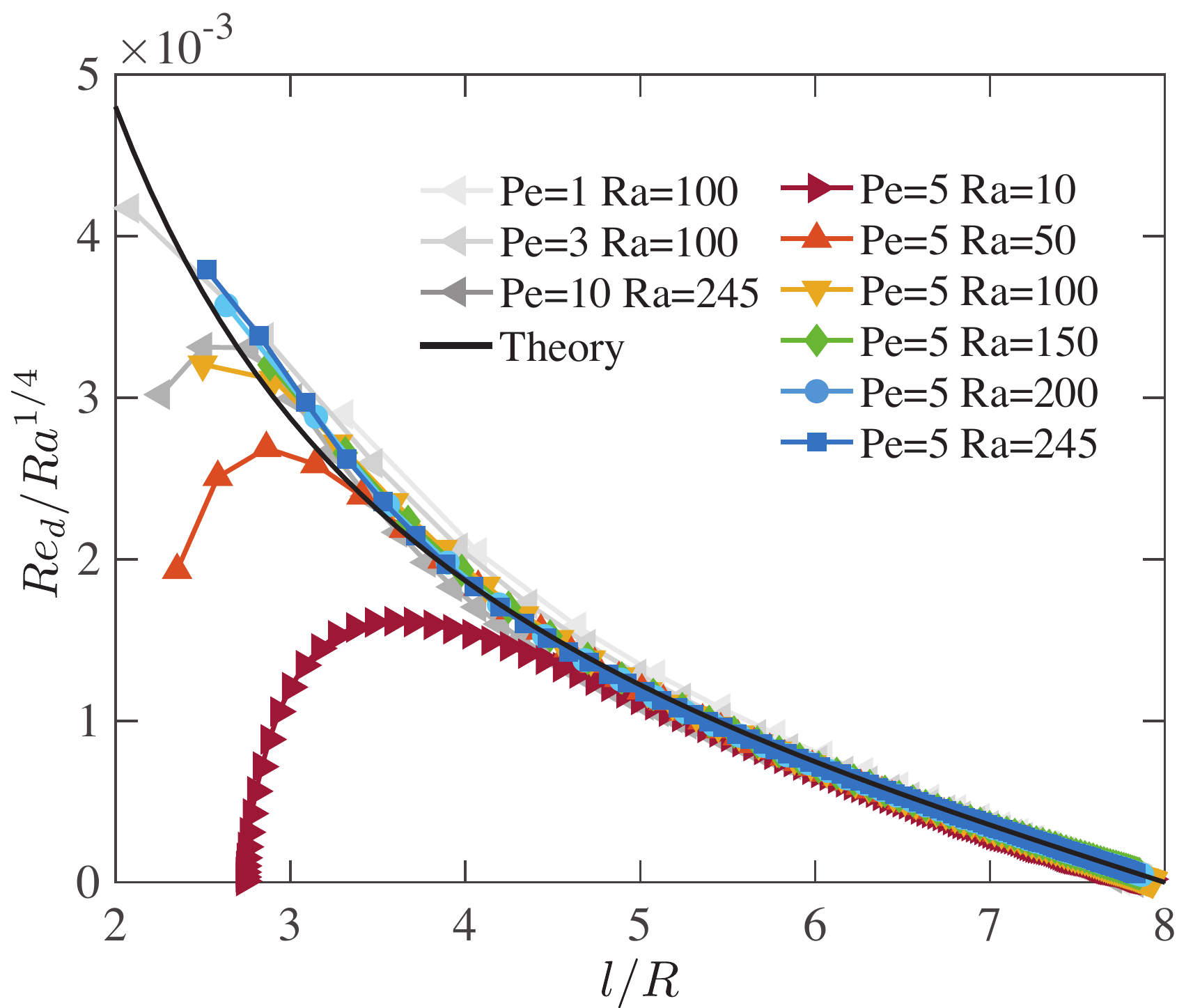}
\caption{$Re_d(\ell)$ normalized by $Ra^{1/4}$ versus the normalized distance $\ell /R$ between the pair of droplets. The markers are for numerical results with lines to guide the eye and the black solid line for relationship (\ref{eqj4_5}) with the fitted prefactor $0.012$.}
\label{fig8}
\end{figure}

\subsection{Droplet velocity using method of reflections and Faxen's law} \label{sec5_2}

In this section, we apply Faxen's law and the method of reflections to account for the interactions between multiple droplets. The principle of the method of reflections is to perform successive approximations for the interaction of droplets within the fluid \citep{guazzelli2011,varma2018}. The velocity of the droplet is calculated iteratively, and in each step, the velocity of the droplet is updated with the disturbance from other droplets using Faxen's law \citep{guazzelli2011}. Despite the far-field assumption of the method, even for close distance $\ell /R\sim O(1)$, it reaches a surprisingly accurate result \citep{ishikawa2006, spagnolie2012hydrodynamics}.

First, we consider a pair of active droplets (droplet 1 and 2) far apart. Since there is no external force and the droplet is isotropic, the droplets are stationary:
\begin{equation}\label{eqja1}
U_1^0=U_2^0=0,
\end{equation} 
where $U_i^j$ represents the velocity of droplet $i$ after the $j$th reflection process.

Then in first reflection, we suppose that the droplets are only moderately far apart, and each droplet makes a disturbance at the velocity of the other. From (\ref{eqj4_3}), droplet 1 causes a fluid velocity disturbance at droplet 2:
\begin{equation}\label{eqja2}
u_2^0\sim \frac{D Ra^{1/4}}{\ell}
\end{equation} 
where $u_i^j$ is the fluid velocity disturbance at the center of droplet $i$ caused by the other droplet after the $j$th reflection.
According to the Faxen's law, the velocity of droplet 2 due to the velocity disturbance caused by droplet 1 is (\cite{guazzelli2011}, p.87):
 \begin{equation}\label{eqja3}
U_2^1= \Big (1+\frac{R^2}{6}\nabla ^2 \Big )u_2^0.
\end{equation} 
Since the two droplets are equivalent, the same velocity is obtained for droplet 1 after the first reflection.

Then we start with the second reflection, the velocity of the droplet obtained in the first reflection will cause disturbance to the other one. The fluid velocity caused by droplet 1 at the center of droplet 2 is (\cite{lamb1924}, p. 599):
\begin{equation}\label{eqja4}
u_2^1= \Big (\frac{3R}{2\ell}-\frac{R^3}{2\ell^3} \Big )U_1^1.
\end{equation} 
Again with Faxen's law, the velocity disturbance of the droplet 2 after the second reflection is given by:
 \begin{equation}\label{eqja5}
U_2^2= \Big (1+\frac{R^2}{6}\nabla ^2 \Big )u_2^1.
\end{equation} 
For higher-order reflection, it is found that the velocity disturbance after reflection
\begin{equation}\label{eqja6}
U_2^n\sim O \Big ( \Big (\frac{R}{\ell} \Big )^{n-1} \Big ).
\end{equation} 
Therefore, we neglect the higher-order small terms, and the velocity of the droplet is approximated as
\begin{equation}\label{eqja7}
U(\ell)=U_2=U_2^0+U_2^1+U_2^2+O \Big (\frac{R}{\ell} \Big )=u_2^0+O \Big (\frac{R}{\ell} \Big )\sim \frac{DRa^{1/4}}{\ell},
\end{equation} 
We define the Reynolds number of the droplet $Re_d$ by the droplet velocity $U$:
\begin{equation}\label{eqja8}
Re_d(\ell)=\frac{UR}{\nu}\sim\frac{D}{\nu}\frac{Ra^{1/4}}{\ell/R}=\frac{1}{Sc}\frac{Ra^{1/4}}{\ell/R}.
\end{equation} 
$Re_d (\ell)$ is different from $Re_y (r)$, where $Re_d(\ell)$ expresses the velocity of a droplet influenced by the other droplet at distance $\ell$, while $Re_y(r)$ corresponds to the fluid velocity at distance $r$ away from a single droplet.

Equation (\ref{eqja8}) considers the influence from only one neighboring droplet. Note that the lateral boundaries are periodic. We consider the influence from the two neighboring droplets and obtain
\begin{equation}\label{eqj4_5}
\frac{Re_d}{Ra^{1/4}}\sim \frac{R}{\ell}-\frac{R}{L_y-\ell}.
\end{equation}

\subsection{Model validation}\label{sec5_3}

The $Re_d$ of the droplets at different distances $\ell$ of different $Ra$ and $Pe$ obtained from simulations are shown in figure \ref{fig8}. The numerical results collapse for sufficiently large droplet separation $\ell$. For large distances $\ell /R$, the numerical results can be described by equation (\ref{eqj4_5}) which excellently agrees with the data for the velocity of the droplet especially for high $Ra$. For low $Ra$ ($Ra\leq50$), there is a deviation between the numerical results and the relationship (\ref{eqj4_5}) near the droplet, which can be explained by the influence of diffusiophoretic flow near the droplet.

\begin{figure}
\centering
\centering \includegraphics[width=1.0\textwidth]{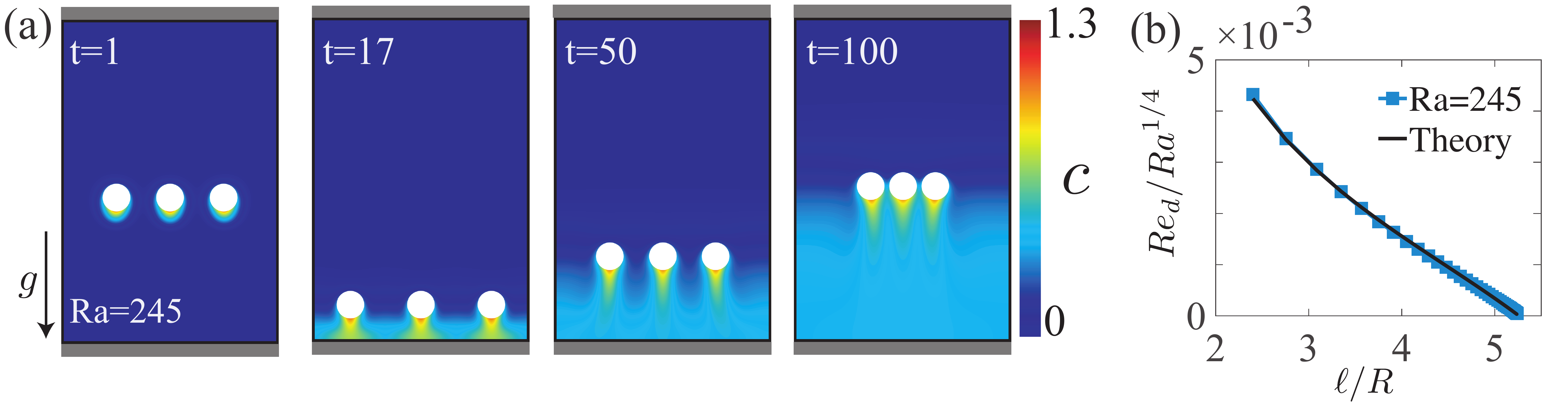}
\caption{(a) Snapshots at different times of concentration fields emerging from three neighboring droplets. Here $Sc=100$, $Pe=5$ and $Ra=245$. (b) $Re_d$ normalized by $Ra^{1/4}$ versus the normalized smallest distance $\ell/R$. The symbols show the numerical results and the solid line shows relationship (\ref{eqj5_1}) with a fitted prefactor $0.013$.}
\label{fig9}
\end{figure}

To further test our theory, we simulate the case of three droplets initially located at the center of the domain with $Ra=245$,  $Sc=100$, and $Pe=5$, where the snapshots are given in figure \ref{fig9}(a). The horizontal velocity of the middle droplet remains at zero due to the symmetry about the middle axis. With our model of subsection \ref{sec5_2}, $Re_d$ follows:
\begin{equation}\label{eqj5_1}
\frac{Re_d}{Ra^{1/4}}\sim \frac{R}{\ell}-\frac{R}{L_y-2\ell}.
\end{equation} 
Indeed, in figure \ref{fig9} (b), for large $\ell /R$ again an excellent agreement is seen between the numerical results and the prediction of equation (\ref{eqj5_1}). 

\begin{figure}
\centering
\centering \includegraphics[width=0.6\textwidth]{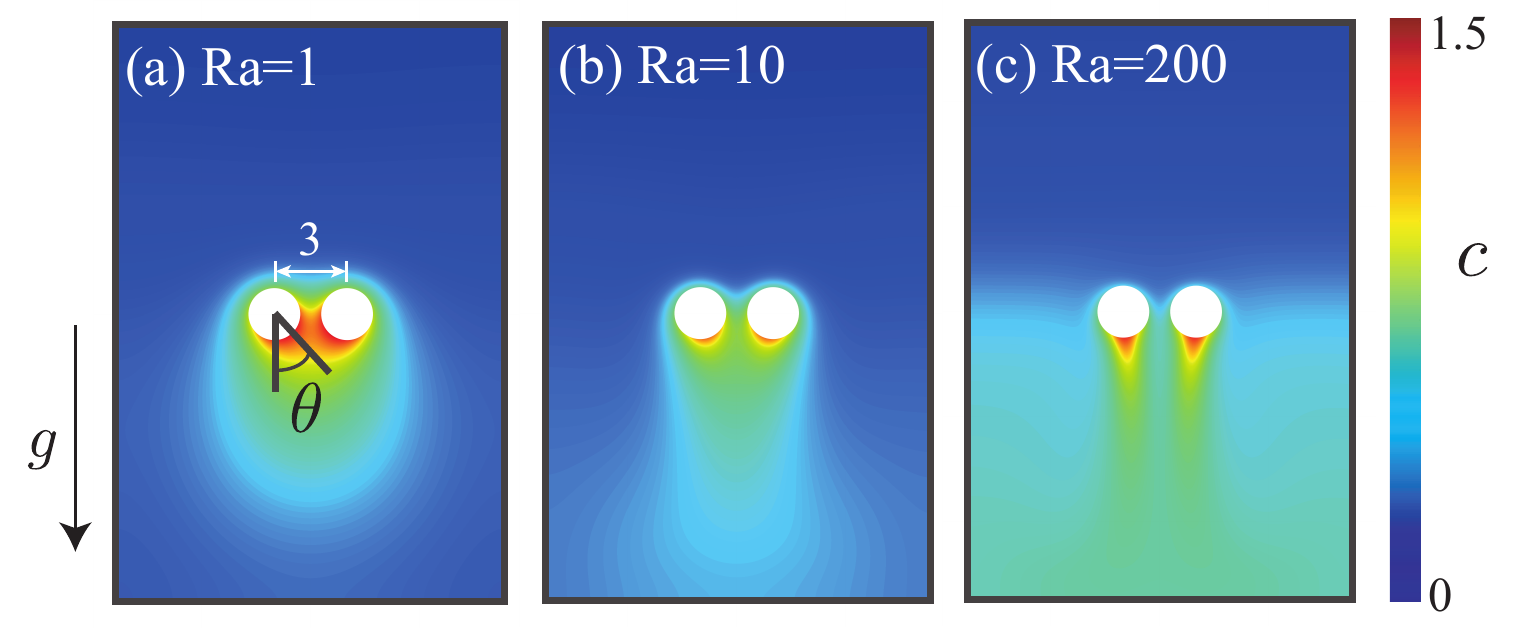}
\caption{The concentration field for a pair of fixed droplets at distance $3$ for $Pe=5$, $Sc=100$ and three different $Ra$ numbers: (a) $Ra=1$, (b) $Ra=10$, (c) $Ra=200$. $\theta$ is the angle between the bottom point and the maximum concentration point to represents the plume position.}
\label{fig12_new}
\end{figure}

\begin{figure}
\centering
\centering \includegraphics[width=0.8\textwidth]{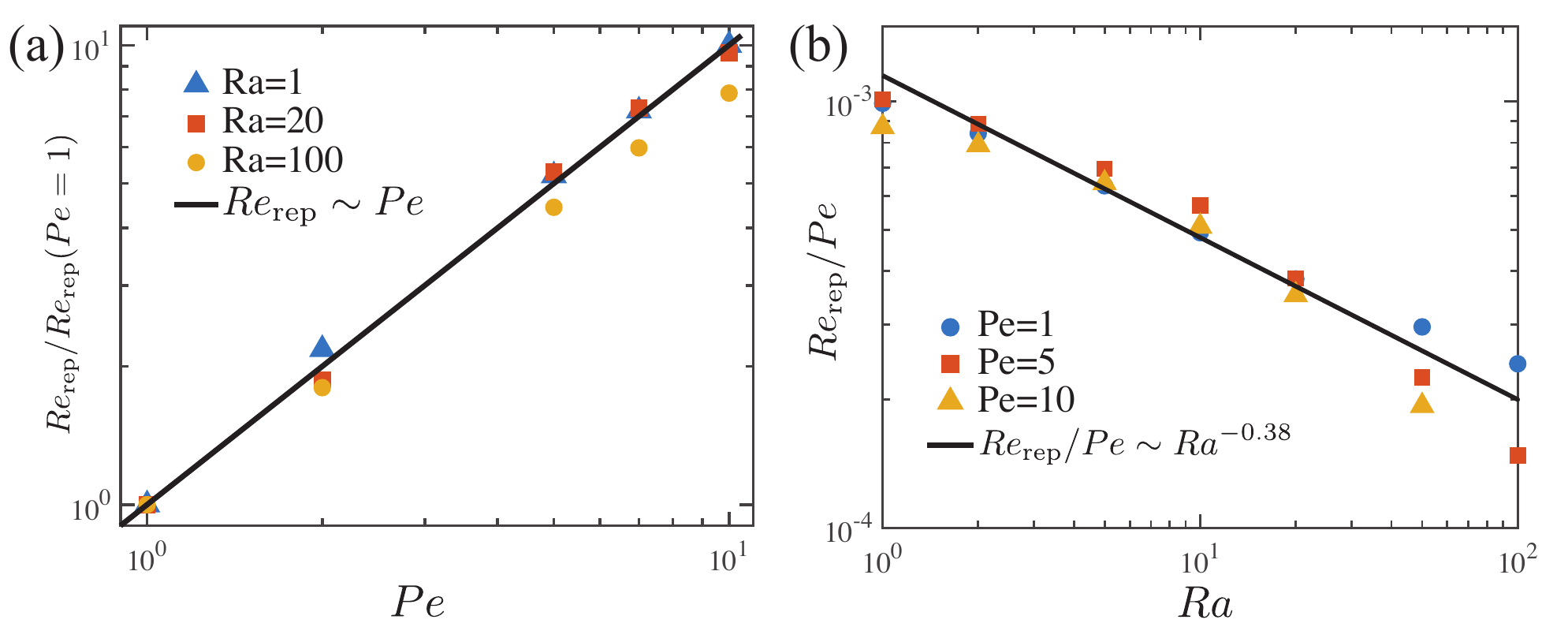}
\caption{(a) Normalized droplet repulsive Reynolds number $Re_{\text{rep}}/Re_{\text{rep}}(Pe=1)$ for different $Pe$. The plot shows that the $Re_{\text{rep}}/Re_{\text{rep}}(Pe=1)$ is proportional to $Pe$.  (b) $Re_{\text{rep}}/Pe$ versus $Ra$. The solid line represents the fitted function, which shows that $Re_{\text{rep}}/Pe$ is proportional to $Ra^{-0.38}$.}
\label{fig13_new}
\end{figure}

\begin{figure}
\centering
\centering \includegraphics[width=0.8\textwidth]{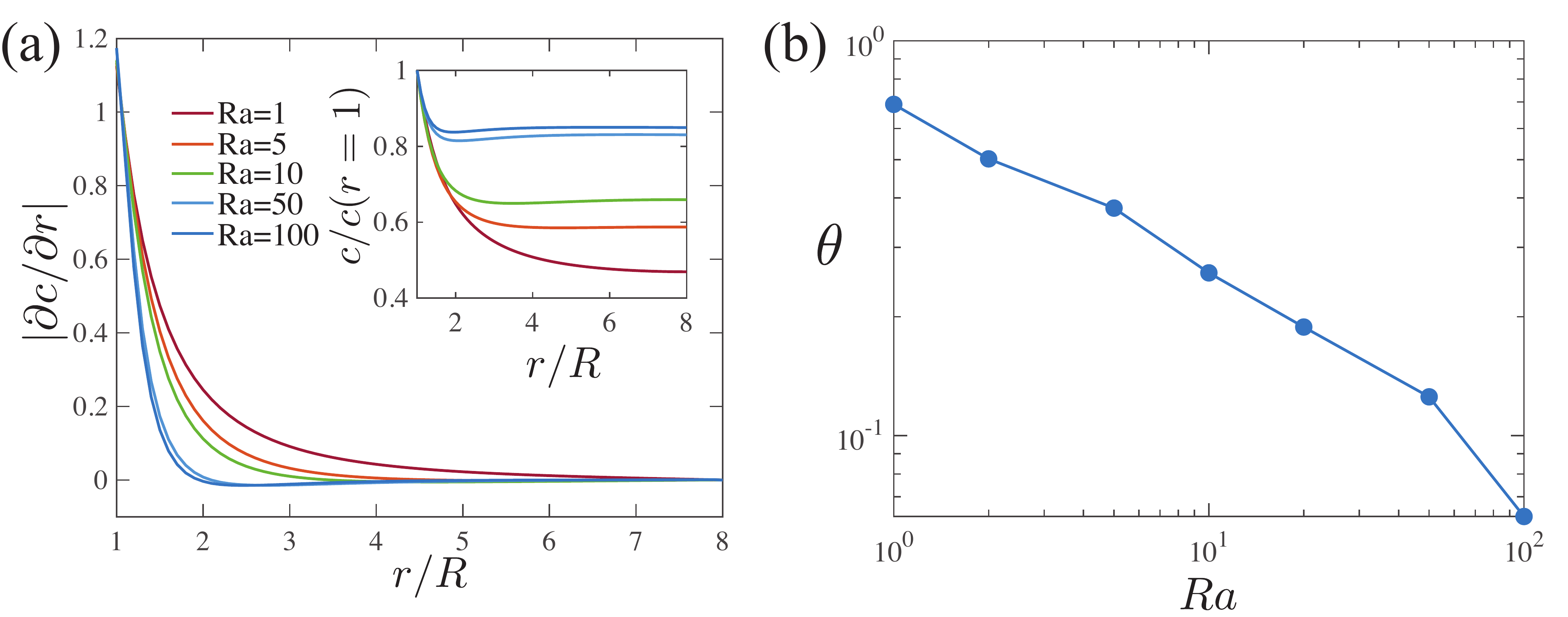}
\caption{(a) Concentration gradient $|\partial c/ \partial r|$ at normlized distance $r/R$ for different $Ra$ near a single droplet, which indicates the concentration gradient decreases as $Ra$ increases. The inset shows the normalized concentration profile. (b) The plume position $\theta$ versus $Ra$, which reflects that the plume is more pulled towards the droplet bottom as $Ra$ increases.}
\label{fig14_new}
\end{figure}

\section{Repulsive effect by diffusiophoresis}\label{sec7}
Given the good agreement between the attraction model with numerical results, now we further study the repulsive velocity from simulations. 

We simulate a pair of droplets fixed at the center of the domain (figure \ref{fig1} (a)) with a horizontal distance $\ell=3$ at $y$ direction. Since the repulsive diffusiophoretic motion mainly comes from the slip velocity induced by concentration field, we estimate the repulsive velocity by the integral of slip velocity at the surface \citep{stone1996propulsion}:
\begin{equation}\label{eqja4}
U_{\text{rep}}=\frac{1}{4\pi R^2}\int_S{\textbf{u}_sdS}.
\end{equation} 
We define $Re_{\text{rep}}=U_{\text{rep}}R/\nu$ to represent the repulsive interaction. We simulate the cases of different $Pe$ and $Ra$, and the resulting concentration field is shown in figure \ref{fig12_new}. The relationship between $Re_{\text{rep}}$ and $Pe$, $Ra$ is shown in figure \ref{fig13_new}. Figure \ref{fig13_new} (a) shows the relationship between the normalized $Re_{\text{rep}}$ and $Pe$. The results indicate that the diffusiophoretic effect leads to repulsive motion, which agrees with our conclusions in Section \ref{sec_result}, and $Re_\text{rep}$ is proportional to $Pe$. Figure \ref{fig13_new} (b) shows $Re_{\text{rep}}$ for different $Ra$, and we find that a stronger buoyancy effect reduces the repulsive velocity between droplets. We fit the results with a power law ansatz and get
\begin{equation}\label{eqja7_2}
Re_{\text{rep}}\sim PeRa^{-0.38}.
\end{equation}

To better understand the decrease of $Re_{\text{rep}}$ with increasing $Ra$, we first study the influence of $Ra$ on its surrounding concentration field. We plot the concentration gradient near a single droplet of different $Ra$ at $Pe=5$ from simulations in Figure \ref{fig14_new} (a).  It indicates that the concentration gradient has a significant drop as $Ra$ increases. This can be rationalized as follows: As $Ra$ increases, buoyancy-driven convection reduces the thickness of the concentration boundary layer \citep{fujii1963,  dietrich2016}. As the surface concentration gradient remains constant (equation (\ref{concengrad})), a reduction in the boundary layer thickness leads to a lower local concentration gradient near the droplet.

Moreover, we find that buoyancy also influences the position of the plume at the droplet surface.  Through the concentration field in figure \ref{fig12_new}, as $Ra$ increases, the plume moves closer to the bottom of the droplet. To evaluate the effect, we define $\theta$ as the angle between the maximum concentration point and the droplet bottom point to represent the position of the plume as indicated in figure \ref{fig12_new}. Figure \ref{fig14_new} (b) shows the change of the plume position for different $Ra$ at $Pe=5$. This finding thus suggests that a stronger buoyancy effect (higher $Ra$) pulls the plume towards the bottom point and this can reduce the horizontal component of the repulsive diffusiophoretic velocity. 

We acknowledge that the arguments above are handwaving and qualitative. The complex system dynamics resulting from the coupling between convection and the concentration field makes a theoretical derivation of the relationship between $Re_{\text{rep}}$ and $Ra$ too challenging.

However, if we combine the equations for the attractive (\ref{eqja8}) and repulsive velocities (\ref{eqja7_2}), we obtain
\begin{equation}\label{eqja7_3}
Pe\sim Ra^{0.63},
\end{equation} 
which perfectly describes the transition between the attracting and the repelling regimes, see figure \ref{fig4} (b). This plot nicely reflects that the mechanism behind the interaction between droplets is the competition between the attractive force by buoyancy and the repulsive force by diffusiophoresis. 

\section{Summary \& Conclusions} \label{con}
We have studied the interaction between droplets with diffusiophoretic effect, droplet buoyancy effect and product buoyancy effect.  The corresponding parameters are P\'eclet number ($Pe$), Galileo number ($Ga$) and Rayleigh number ($Ra$). We have simulated the cases over a range of $Pe$, $Ra$, and $Ga$, with $Sc$ being fixed at $100$. 

For a pair of droplets, using numerical simulations, we have found that the product buoyancy effect leads to the attractive motion between droplets, while the Marangoni/diffusiophoretic effect results in repulsion. A larger $Pe$ results in a larger $Ra$ threshold, above which droplet collision occurs. If the Rayleigh number is sufficiently small, the distance between droplets reaches an equi-distance equilibrium, and as $Ra$ increases, the closest balanced distance decreases, which indicates that the product buoyancy weakens the repulsion caused by the Marangoni/diffusiophoretic effect. For sufficiently high Rayleigh numbers ($Ra\geq50$), the droplets collide with each other. Then we investigated the influence of droplet buoyancy effect and found that the attracting behavior is similar for different $Ga$, and the change of $Ga$ only leads to a delayed collision.

With the simulation of a single droplet, we have found that the attraction originates from convective flow induced by the density difference between the dissolving product and ambient fluid. Based on this, we have created a simple model which well describes the horizontal velocity near the droplet. The local Reynolds number is inversely proportional to the distance from the droplet as shown in equation (\ref{eqj4_3}).

With the above model as a starting point, we have obtained the equation for the attracting velocity of the droplet at high $Ra$ with Faxen's law and the method of reflections. The attracting velocity is proportional to $Ra^{1/4}$ and inversely proportional to the distance between the droplets. The results have been verified by the simulation results for cases with two and three droplets.

Then we have investigated the repulsive effect by simulating the case of a pair of fixed droplets and the repulsive velocity was approximated by the integral of the slip velocity (\ref{eqja4}). We have found that $Re_{\text{rep}}$, which represents the repulsive velocity, is proportional to $PeRa^{-0.38}$. The linear dependence of $Re_{\text{rep}}$ on $Pe$ is simply due to a larger diffusiophoretic repulsive force for larger $Pe$. In contrast, the $Ra$-dependence of $Re_{\text{rep}}$ is more complicated. It reflects that an increasing $Ra$ leads to a smaller horizontal concentration gradient and favours the plume to be closer to the bottom point of droplet, which reduces the repulsive velocity. 

Combining the scaling relations of the attractive and repulsive velocity, we obtain $Pe\sim Ra^{0.63}$, which perfectly describes the transition curve between the attractive and repulsive regime in figure \ref{fig4} (b). This indicates that the mechanism behind the interaction between droplets are the competition between attractive buoyancy force and repulsive diffusiophoretic force.

The present work contributes to the understanding of the interaction between active droplets, and specifically reveals the significant role played by the dissolving product buoyancy. It shows that product buoyancy can lead to attractive motion between active particles, which helps us understand the attraction of active droplet in the experiments of  \cite{krueger2016}. We have proposed a simple model to predict the velocity of the interacting active droplets. The results provide a framework to understand the droplet attraction induced by the convective flow. Moreover, the present work reveals a possible way to change the collective behaviors by tuning the buoyancy.

In our simulation, the propulsion of the active droplet is simply modelled as diffusiophoreis. Alternatively, we could have taken Marangoni flow. Until now, buoyancy-driven attractive motions are only observed in the cases of active droplets but scarcely in phoretic particles, possibly due to the difficulties to generate large enough density difference between the product and ambient fluid by phoretic particles.

Many questions remain open. For example, how to determine the cluster size for multiple droplets? How does the flow field change if droplets are near a fluid-air interface? How to quantitatively determine the threshold Rayleigh number above which the droplets collide with each other and show collective behaviours? With the obtained insights into the attraction here, we hope it is seen as worthwhile to further explore the formation and motion of a cluster of active particles.

\section*{Acknowledgements}
We greatly appreciate valuable discussions with Martin Assen, Utkarsh Jain, and Corinna Maass. We acknowledge the support from the Netherlands Center for Multiscale Catalytic Energy Conversion (MCEC), an NWO Gravitation program funded by the Ministry of Education and support from the ERC-Advanced Grant "DDD" under the project number 740479. The simulations in this work were carried out on the national e-infrastructure of SURFsara, a subsidiary of SURF cooperation, the collaborative ICT organization for Dutch education and research, MareNostrum 4 which is based in Spain at the Barcelona Computing Center (BSC) under PRACE projects 2018194742, 2020225335 and 2020235589, on Irene at Trés Grand Centre de calcul du CEA (TGCC) under PRACE project 2019215098 and on Marconi successor at CINECA, Italy under PRACE project 2019204979. K. L. Chong is supported by the Natural Science Foundation of China under Grant No. 92052201. 

\section*{Declaration of interests}
The authors report no conflict of interest.

\bibliographystyle{jfm}

\bibliography{reference_catalyticplane}

\end{document}